\documentclass[apj]{emulateapj}

\usepackage{amsmath}	% Advanced maths commands
\usepackage{amssymb}	% Extra maths symbols

\newcommand{\Msun}{\ensuremath{M_\odot}}

\slugcomment{submitted to Ap. J. (July 26, 2018)}

\shorttitle{Minimum Period of Pre-CVs}
\shortauthors{Nelson et al.}

\begin{document}

\title{Minimum Orbital Period of Pre-Cataclysmic Variables}

\author{L. Nelson}%\altaffilmark{1}}
\affil{Department of Physics \& Astronomy, Bishop's University, Sherbrooke, QC J1M 1Z7}
    \email{{\it Corresponding author}: lnelson@ubishops.ca}

\author{J. Schwab\altaffilmark{1}}
\affil{Department of Astronomy \& Astrophysics, University of California, Santa Cruz, CA 95064}

\author{M. Ristic}
\affil{Department of Physics, Northeastern University, Boston, MA 02115}
\and

\author{S. Rappaport}
\affil{Department of Physics, and Kavli Institute for Astrophysics \& Space Research, M.I.T., Cambridge, MA 02139, USA}

\altaffiltext{1}{Hubble Fellow}

\begin{abstract}
More than 20 pre-cataclysmic variable (pre-CV) systems have now been discovered with very short orbital periods ranging from 250 min down to 68 min.  A pre-CV consists of a white dwarf or hot subdwarf primary and a low-mass companion star, where the companion star has successfully ejected the common envelope of the primary progenitor, but mass transfer from the companion star to the primary has not yet commenced.  In this short-period range, a substantial fraction of the companion stars are likely to be either brown dwarfs with masses $\lesssim 0.07 \, M_\odot$ or stars at the bottom of the MS  ($\lesssim 0.1 M_\odot$).   The discovery of these short-period pre-CVs raises the question -- what is the shortest possible orbital period of such systems?   We ran 500 brown dwarf/low-mass main sequence models with {\tt MESA} that cover the mass range from 0.002 to 0.1 $M_\odot$.  We find the shortest possible orbital period is 40 min with a corresponding brown dwarf mass of 0.07 $M_\odot$ for an age equal to a Hubble time.  We discuss the past evolution of these systems through the common envelope and suggest that many of the systems with present day white dwarf primaries may have exited the common envelope with the primary as a helium burning hot subdwarf.  We also characterize the future evolution of the observed systems, which includes a phase as CVs below the conventional period minimum.
\end{abstract}


\section{Introduction}

More than 20 short-period, detached binary systems consisting of a white dwarf (WD) or hot subdwarf (sdB) primary and an ultra-low mass companion ($\lesssim 0.1 M_\odot$) have been discovered during the past decade.  Although it is not impossible that some of these companions descended from highly evolved stars that experienced mass exchange with the primary during a previous epoch, this scenario is considered highly unlikely based on the available observational evidence.  Instead we expect that most, if not all, of these companions are either stars at the bottom of the main sequence (MS), {\it bona fide}  brown dwarfs (BDs) with masses of $\lesssim 0.072 \, M_\odot$, or `transition objects' with masses in the range of $0.072 \lesssim \, M/M_\odot \lesssim 0.075$ (all of solar metallicity).  It is reasonable to assume that these ultra-low mass companions were born in binaries with low- and intermediate-mass stars that ultimately evolved to become giants with either helium (He) or carbon-oxygen (CO) cores.  If the separation of the progenitor binary were sufficiently close, the BD would be engulfed by the expanding giant leading to dynamical instability. During the subsequent evolution, it is expected that the BD would expel the giant's hydrogen-rich envelope as it spirals in towards the core.  If a merger can be avoided, a very tight binary consisting of the relatively unperturbed BD and a compact companion is produced.  This process is known as common envelope (CE) evolution and it takes place on a short timescale of less than 1000 years (see, e.g., Xiong et al. 2017, and references therein).  The known short period ($P \lesssim 4$ hr) pre-CVs with ultra-low mass ($M \lesssim 0.1\,M_\odot$) companions are summarized in Table \ref{tbl:preCVs}.

\begin{table*}
\centering
\caption{White Dwarfs or Hot Subdwarfs in Close Detached Binaries with Very Low Mass ($\la 0.1\,M_\odot$) Companions}
\begin{tabular}{lccccccc}
\hline
\hline
 Object Name & $P_{\rm orb}$ & $M_{\rm primary}$ & $T_{\rm eff}$ & $M_{\rm companion}$ & Eclipsing & Reference  \\
  & (min) & ($M_\odot$) & (K) & ($M_\odot$) &  \\
\hline
%GD 1400 & 600  &  0.67   &  11,600  &  0.06$^a$  & no &  1,2,3  \\
WD 0837+185 & 250  &  $0.798\pm0.006$  &  15,000  &  $\ge 0.024$$^b$  & no & Casewell et al.~(2012) \\
2MASS J15334944+3759282$^{a}$ & 233 & $0.376 \pm 0.055$ & 29,230 & $0.113 \pm 0.017$ & yes & For et al.~(2010) \\
NN Ser & 187 & $0.535 \pm 0.012$ & 57,000 & $0.111 \pm 0.004$ & yes & Parsons et al.~(2010) \\
% 2MASS J19383260+4603591$^{a}$ & 181& $0.372 \pm 0.024$ & 29,564  & $0.1002 \pm 0.0065$ & yes & Barlow et al.~(2012) \\
2MASS J19383260+4603591$^{a,c}$ & 181& $0.48 \pm 0.03$ & 29,564  & $0.12 \pm 0.01$ & yes & {\O}stensen et al.~(2010) \\
EC 10246-2707$^{a}$& 171& $0.45 \pm 0.17$ & 28,900  & $0.12 \pm 0.05$ & yes & Barlow et al.~(2013) \\
SDSS J135523.91+085645.4 & 165 & $0.46 \pm 0.01$ & 33,160 & $0.090 \pm 0.007$ & no & Badenes et al.~(2013) \\
HS 2231+2441$^{a,d}$ & 159 & $0.265 \pm 0.010$$^e$ & $28,500$ & $\simeq 0.05 $ & yes & Ostensen et al.~(2008) \\  % \pm 80
NSVS 14256825$^a$ & 159 & $0.346 \pm 0.079$ & $42,300$ & $0.097 \pm 0.028$ & yes & Almeida et al.~(2012), Model 1  \\  % \pm 400
UVEX J032855.25+503529.8$^{a}$ & 159 & $0.49 \pm 0.050$ &  28,500  & $0.120 \pm 0.010$ & yes &  Kupfer et al.~(2014)\\ 
PTF1 J085713+331843 & 153 & $0.61\pm0.18$ & $25,000$ & $0.19\pm0.10$ & yes & van Roestel et al.~(2017) \\
GD 488  & 148 & $0.41\pm0.01$ & $19,600$ & $0.096\pm0.004$ & no & Maxted et al.~(1998) \\
PG 1336-018$^{a}$   & 145 & $0.466\pm0.006$ & 32,400 & $0.122\pm0.001$ & yes & Vu\v{c}kovi\'{c} et al.~(2007), Model II \\
SDSS J082053.53+000843.4$^a$ & 139 & $\simeq 0.47$$^{f}$ & $26,700$ & $0.068 \pm 0.003$ & yes & Geier et al.~(2011c) \\ %  \pm 100
HS 0705+6700$^a$ & 138 & $\simeq 0.48$ & $28,800$ & $\simeq 0.13$ & yes &Drechsel et al.~(2001) \\ 
SDSS J155720.77+091624.6 & 136 & $0.447\pm0.043$ & 21,800 & $0.063\pm0.002$ & no & Farihi et al.~(2017) \\
SDSS J141126.20+200911.1 & 122 & $0.53 \pm 0.03$ & 13,000 & $0.050 \pm 0.002$ & yes & Littlefair et al.~(2014) \\
WD 0137-349 & 116  &  $0.39\pm0.035$   &  16,500  &  $0.053 \pm 0.006$   & no & Maxted et al.~(2006)$^{g}$ \\
PG 1017-086$^{a}$  & 105  &  $\simeq 0.5^{f}$   &  30,300  &  $0.078 \pm 0.006$   & yes & Maxted et al.~(2002) \\
NLTT 5306 & 102   &  $0.44 \pm 0.04$ & 7,756  & $0.053 \pm 0.003$ & no & Steele et al.~(2013)  \\
SDSS J162256.66+473051.1$^a$ & 100 & $0.48 \pm 0.03$ & $29,000$ & $0.064 \pm 0.004$  & yes & Schaffenroth et al.~(2014) \\ %  \pm 600
V2008-1753$^a$ & 94.8 & $0.47 \pm 0.03$ & $32,800$ & $0.069 \pm 0.005$ & yes & Schaffenroth et al.~(2015) \\ %  \pm 750
SDSS J085746.18+034255.3 & 93.7 & $0.514 \pm 0.049$ & $35,300$ & $0.087 \pm 0.012$ & yes & Parsons et al.~(2012) \\ 
SDSS J123127.14+004132.9 & 72.5 & $0.56 \pm 0.07$ & $37,210$ & $\lesssim 0.095$ & yes & Parsons et al. (2017) \\ %  \pm 1140
WD 1202-024 & 71.2  &  $0.40 \pm 0.02$  &  $22,650$   &  $0.055 \pm 0.008$  & yes & Rappaport et al.~(2017)  \\  %  \pm 540
EPIC 212235321 & 68.2  &  $0.47 \pm 0.01$  &  $24,490$   &  $\simeq 0.063$  & no & Casewell et al.~(2018) \\  %  \pm 150
\hline
\label{tbl:preCVs}
\end{tabular}

{\bf Notes.} This table draws upon previous compilations by Ritter \& Kolb (2003; RKcat 7th ed., V7.21), the MUCHFUSS project (Kupfer et al. 2015, Schaffenroth et al. 2018), and Parsons et al. (2015).  (a) Primary is a hot subdwarf.  (b) This value is $M \sin i$.
(c) Barlow et al. (2012) report $M_{\rm primary} = 0.372 \pm 0.024$ \Msun\ and $M_{\rm companion} = 0.1002 \pm 0.0065$ \Msun.
(d) Almeida et al. (2017) also report solutions with small primary masses: $M_{\rm primary}, M_{\rm companion}$ = 0.19 \Msun, 0.036 \Msun\ or 0.288 \Msun, 0.046 \Msun.
(e) This object is too low in mass to be a He-core-burning star.  (f) This model assumes a canonical hot subdwarf mass.  (g) Also see Burleigh et al. (2006) for more details.
\end{table*}

Systems with BD companions are of particular interest because: (i) their small radius permits short orbital periods leading to potentially large fluxes of gravitational radiation that might be detectable with eLISA; and, (ii) their low masses make them ideal probes of one of the extremes of CE evolution.  These WD+BD and sdB+BD binaries are taken to represent the pre-cataclysmic variable (pre-CV) phase of systems that will become mass-transferring cataclysmic variables (CVs) once the orbit shrinks to the point where the BD fills its Roche lobe{\footnote{Under the assumption that the orbit shrinks at a rate set by gravitational wave radiation, the sdB primaries will have evolved to become WDs by the time mass transfer begins (see Section 2).}}.  For a conventional (unevolved) CV  system, a low-mass star ($\lesssim 1 M_\odot$) is brought into contact with the WD accretor as a result of orbital decay due to angular momentum losses from gravitational radiation and/or magnetic braking (see, e.g., Paczy\'{n}ski \& Sienkiewicz 1981; Rappaport, Joss, \& Webbink 1982; Nelson \& Goliasch 2015, and references therein).  Once the donor overfills its Roche lobe, mass transfer drives the orbital evolution from periods of several hours down to an observed orbital period minimum ($P_{\rm CV,min}$) of about 80 minutes.  The typical donor-star mass at this juncture is $\approx 0.06 \, M_\odot$ and the donor continues to lose mass with a concomitant increase in the period.  

The pre-CVs containing the ultra-low mass companions listed in Table \ref{tbl:preCVs} will also be forced into contact largely as a result of gravitational radiation losses.  However, when these systems start mass transfer they will have orbital periods shorter than $P_{\rm CV,min}$ even though the masses of the donors are comparable.  The reason is that the donors in conventional CVs at $P_{\rm CV,min}$ are in the process of losing mass and this causes them to depart from thermal equilibrium.  The `thermal bloating' that they experience causes them to be considerably larger in radius than the low-mass companions in pre-CVs.  Thus when the pre-CV companions overfill their Roche lobes they do so at considerably shorter periods.

With the ever increasing number of short-period, ultra-low mass, pre-CV discoveries being made, this channel for producing CVs has attracted considerable attention.   Politano (2004) and Politano \& Weiler (2007) investigated this channel by carrying out a  comprehensive population synthesis analysis.  Detailed analyses of post common envelope binary (PCEB) evolution leading to the formation of pre-CVs was also  undertaken by Davis et al. (2008, 2010), Zorotovic et al. (2010), and Zorotovic \& Schreiber (2012).  Understanding the CE phase of binary evolution is extremely complex and the observations are often used to constrain the models.  Recent discoveries include \mbox{WD 1202-024} (Rappaport et al. 2017; Parsons et al.  2017) with $P_{\rm orb}=71.2$ minutes and EPIC 212235321 (Casewell et al. 2018) with $P_{\rm orb}=68.2$ minutes.  In the former case, Rappaport et al. (2017) showed that this binary would become a CV in about 250 Myr with a $P_{\rm orb}\simeq 55$ minutes (much less than $P_{\rm CV,min}$).  The first CV thought to have been possibly formed via this channel is SDSS  J150722.30+523039.8 (Littlefair et al., 2007).  Littlefair et al. argued that based on: (i) the short orbital period of only 66.6 minutes (well below $P_{\rm CV,min}$); and, (ii) the lack of helium observed in its spectrum (implying that the donor is not chemically evolved), the progenitor of J1507 was quite plausibly a BD+WD pre-CV binary. 

Given the observational evidence and the fact that these pre-CVs are being found at ever shorter orbital periods leads us to consider the question -- what are the shortest possible periods for such systems?  Moreover, since these systems must have a different evolutionary history than conventional CVs, it is important to understand their formation and subsequent evolution.  Presumably the pre-CVs with BD companions emerged directly from a CE phase that ejected the envelope of the primary progenitor star.  These systems then provide interesting constraints on the parameters associated with the common envelope process.  This raises a related question -- what are the lowest mass BDs that would be able to successfully eject the common envelope of the WD progenitor star?

The main goal of this paper is to address these two questions.  In Section~\ref{sec:formation} we briefly review the common-envelope formation scenario and derive a relation for how the post-common envelope period of the binary depends on the mass of the two stars and mass of the primary progenitor.  In Section~\ref{sec:pmin1} we derive a relation for the minimum period of a pre-CV binary before Roche lobe overflow commences.  Section~\ref{sec:MESA} presents a sequence of BD models generated with the MESA stellar evolution code.  In Section~\ref{sec:pmin2} we utilize the MESA models to find the minimum orbital period of these pre-CV binaries as a function of the BD cooling age.  Section~\ref{sec:subeq_evol} describes the evolution of these systems after mass transfer begins, and Section~\ref{sec:minmass} addresses the question of the minimum companion mass necessary to eject the common envelope.  In Section \ref{sec:popsyn} we describe a rudimentary population synthesis study to evaluate where systems might be expected to be found in the orbital period--brown dwarf mass plane.   Section~\ref{sec:conclusion} contains a summary and our conclusions.

\section{Formation Scenario}
\label{sec:formation}

\begin{figure}
\begin{center}
\includegraphics[width=\columnwidth]{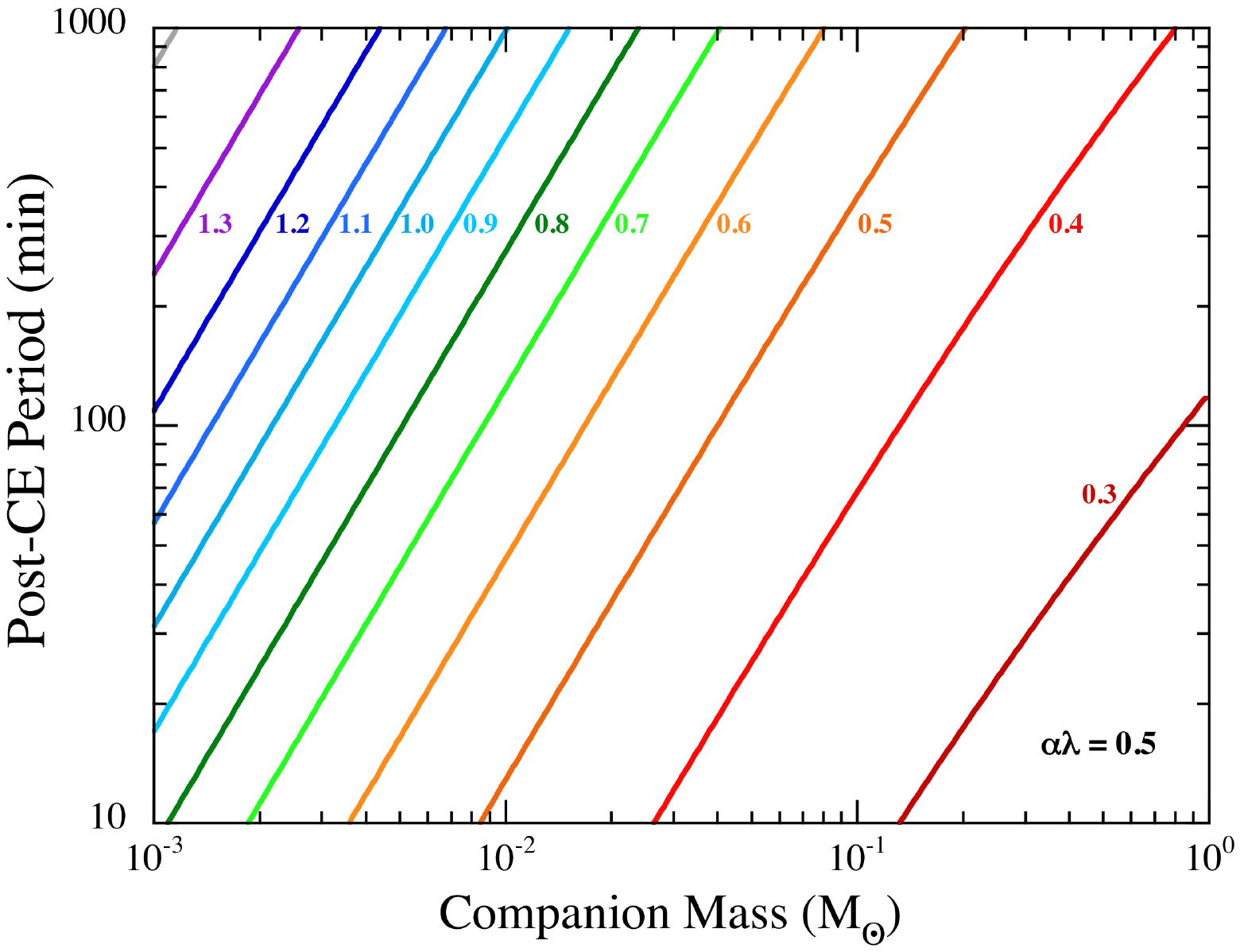}
\caption{Post common-envelope orbital periods as a function of companion mass.  The colored curves and labels are for different WD or sdB masses, in $M_\odot$. The progenitor mass was fixed at an illustrative value of 1.5 $M_\odot$.  The energy efficiency factor for ejecting the common envelope, $\alpha \lambda$, was taken to be 0.5.  These curves are without regard as to whether the companion would be overflowing its Roche lobe at the end of the CE phase (this issue is addressed later in Sects.~\ref{sec:pmin2} and \ref{sec:minmass}).
}
\label{fig:pcep} % Figure 1
\end{center}
\end{figure}

In our preferred formation scenario for these pre-CVs, the giant in the primordial binary undergoes a common-envelope phase wherein the BD strips off the envelope of the giant thereby unveiling its hot core.  To explore this scenario more quantitatively, we invoke the $\alpha$-formulation that uses energy considerations to determine how binaries experiencing a CE phase will evolve (see, e.g., Paczy\'nski 1976; Webbink 1984; Pfahl et al. 2003, and references therein){\footnote{We also note that the $\gamma$-formulation (Nelemans \& Tout 2005) that takes into account angular momentum conservation could also be employed but its physical motivation has been called into question (Woods et al., 2012)}.  Although there are a number of different formulations of the energy-based analysis, the ultimate goal is to determine the final binary orbital separation, $a_f$, once the CE has been ejected, in terms of the initial orbital separation of the primordial binary, $a_i$, and its component masses.  The more complex representations of the energy formulation take into account the fraction of the internal energy used to eject the envelope, for example the recombination energy (see, e.g., Zorotovic et al. 2010).  We have elected to employ a simpler parameterization that reduces the number of free parameters.  Following de Kool (1990), we take 
\begin{equation}
\frac{GM_p M_e}{\lambda r_L a_i}=\alpha \left[\frac{GM_c M_s}{2 a_f}-\frac{G M_p M_s}{2a_i}\right],
\label{eqn:CE1}
\end{equation}
where $M_p$ and $M_s$ are the masses of the primordial primary (the sdB or WD progenitor) and the primordial secondary star (the companion BD), respectively, and $M_c$ and $M_e$ are the masses of the core and envelope of the primary star (see, e.g., Taam et al. 1978; Webbink 1984; Taam \& Bodenheimer 1992). The parameter $\lambda^{-1}$ is a measure of the total binding energy of the envelope to the core of the primary star in units of $-GM_p M_e/R_p$, while $\alpha$ is an energy efficiency parameter for ejecting the common envelope.  The factor $r_L \equiv R_L/a_i$ is the dimensionless radius of the Roche lobe of the primary star when mass transfer starts.  

For the stellar masses and separations involved in the formation of these pre-CV binaries, the second term in square brackets in Eqn.~(\ref{eqn:CE1}) is negligible compared to the first term (see Rappaport et al.~2015 for a more detailed analysis).  Dropping that term, we find:
\begin{equation}
\frac{a_f}{a_i} \simeq \frac{\lambda \alpha r_L}{2} \left(\frac{M_c M_s}{M_e M_p}\right).
\label{eqn:CE2}
\end{equation}
The ratio of final to initial orbital periods follows from Kepler's third law:
\begin{equation}
\frac{P_f}{P_i} \simeq \left(\frac{\lambda \alpha r_L}{2}\right)^{3/2} \left(\frac{M_c M_s}{M_e M_p}\right)^{3/2} \left(\frac{M_p+M_s}{M_c+M_s}\right)^{1/2}.
\label{eqn:CE3}
\end{equation}

We can go one step further with Eqn.~(\ref{eqn:CE3}) and eliminate the initial orbital period, $P_i$, just prior to the start of the common-envelope phase, in favor of an expression that relates $P_i$ to $M_c$, $M_p$, and $M_s$.  Here we utilize Eqn.~(7) of Rappaport et al.~(1995) which has been here generalized to allow for a non-zero envelope mass of the giant:
\begin{equation}
P_i  \simeq  \frac{4 \times 10^4 \, m_{\rm c}^{6.75}}{(1+4 m_{\rm c}^4)^{3/2}} \frac{1}{r_L^{3/2}}\frac{1}{\sqrt{m_p+m_s}}~~{\rm days}
\label{eqn:PM}
\end{equation}
(see also Tauris \& van den Heuvel 2014) where the lower-case masses are expressed in solar units.  Here $r_L$ has the same meaning as in Eqns.~(\ref{eqn:CE2}) and (\ref{eqn:CE3}).  Note that Eqn.~(\ref{eqn:PM}) is based on the nearly unique relation between the core mass of a low-mass giant and its radius (see Eqn.~(5) of Rappaport et al.~1995), applies to both the red giant branch (RGB) and asymptotic giant branch (AGB), and is valid up to a primary mass of $\approx 2.5\,\Msun$.

We now combine Eqns.~(\ref{eqn:CE3}) and (\ref{eqn:PM}) into a single equation for the post-common envelope period and associate the system masses in Eqn.~(\ref{eqn:CE3}) with those we observe in the pre-CV binaries: $M_c \equiv M_{\rm wd}$, $M_s \equiv M_{\rm bd}$, and $M_e \equiv M_p - M_c$.  This yields  
\begin{eqnarray}
P_{\rm PCE} & \simeq & \left(\frac{\lambda \alpha }{2}\right)^{3/2}  \frac{4 \times 10^4 \, m_{\rm wd}^{6.75}}{(1+4 m_{\rm wd}^4)^{3/2}} ~~ \times \nonumber \\
&  & \left[\frac{m_{\rm wd}\,m_{\rm bd}}{(m_p-m_{\rm wd}) m_p}\right]^{3/2}  \left(m_{\rm wd}+m_{\rm bd}\right)^{-1/2}  ~~{\rm days}. \nonumber \\
\label{eqn:WDBD}
\end{eqnarray}
For notational convenience, we use the subscript ``wd'' to represent the present day primary, with the understanding that in some systems the primary is presently an sdB star (see Table 1); likewise, we use ``bd'' to represent the present day secondary, even through some are likely MS stars. Note that the period of the post-CE pre-CV binary is a function only of the masses of the BD and the WD (or sdB) and its progenitor.

We show in Fig.~\ref{fig:pcep} a plot of the period of the post-CE pre-CV binaries as a function of the BD mass for a wide range of discrete values of the WD mass.  In all cases, we took an illustrative value of 1.5 $M_\odot$ for the mass of the WD progenitor star.  We also adopted a conservative value of $\alpha \lambda = 0.5$; smaller values lead to even shorter post-CE periods.  We will discuss how these relationships can be used to infer a minimum companion mass in Section~\ref{sec:minmass}.

We note that the high mass of WD 0837+185 (Table \ref{tbl:preCVs}) implies that it is a CO WD and thus that this system underwent its CE on the AGB.  However, the majority of the other pre-CV systems have primary masses $\la 0.47\, M_\odot$, consistent with the CE occurring on the RGB.  If this interaction occurs such that the He core of the primary has not reached He-ignition, then this leaves behind a He WD primary.  (One of the observed systems, HS2231+2441, has a primary that is too low in mass to be a He-core-burning star, strongly implying that it is a young He WD produced in this way.)  If the He core of the primary does reach He ignition, then it leaves behind a He-core-burning primary that will likely evolve to become an sdB star.  If the post-CE period is not so short that the system comes into contact during the He-burning lifetime of the sdB star, then by the onset of mass transfer the primary will be a low mass CO-core WD and the system will also spend time as a detached WD+BD binary.  As suggested in Schaffenroth et al. (2018), the fact that the WD masses in the known WD+BD binaries are $\approx 0.5 \Msun$ means it is possible that most of these systems could have emerged from the CE as sdB+BD binaries.

In Figure~\ref{fig:contact} we show the timescale for the observed
systems from Table~\ref{tbl:preCVs} to reach contact, assuming that
the orbit is evolving only due to gravitational waves (Peters 1964).
To determine the size of the orbit at contact, we assume that the
secondary has the radius of a model with an age of 10 Gyr; we discuss
how these radii are calculated in Sections~\ref{sec:MESA} and
\ref{sec:pmin2}. The grey line indicates the characteristic He-burning
lifetime ($\approx 150$ Myr) of a canonical mass sdB (e.g.,
Schindler et al. 2015).  All of the systems with sdB primaries have
inspiral times longer than this, meaning the sdB will have become a WD
by the time the system reaches contact.  (This was pointed out for the MUCHFUSS sample by Kupfer et al. 2015; see their Table 5.)
This indicates that the
observed sdB+BD systems can be the direct progenitors of the WD+BD
systems.  For example, the 68 min system (EPIC 212235321) has a WD of 0.47 $M_\odot$
with a cooling age of only $\approx 18$ Myr (Casewell et al. 2018).  A
plausible history for this system is that it emerged from the CE with an orbital 
period of $\approx 80$ min, appeared as an HW Vir system\footnote{This class of eclipsing binary consists of a low-mass MS star or BD in a short-period orbit with a subdwarf B (sdB) star.} as the period shrank
to $\approx 70$ min, at which point the WD formed, and then after a
short phase of WD cooling and further inspiral, the binary reached its
current state.

\begin{figure}
\begin{center}
\includegraphics[width=\columnwidth]{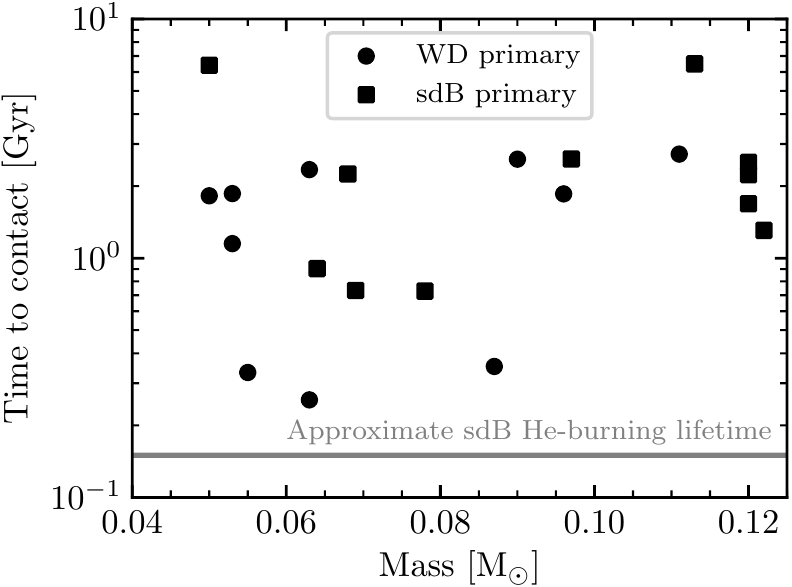}
\caption{Timescale for gravitational waves to bring the observed systems into contact.  Systems with WD primaries are shown as circles; systems with sdB primaries are shown as squares. A characteristic sdB He-burning lifetime is shown as the gray horizontal line.}
\label{fig:contact}
\end{center}
\end{figure} % Figure 2

\section{Dependence of $P_{\rm min}$ on Mass and Radius}
\label{sec:pmin1}

If the low-mass companion star is able to successfully eject the envelope of the progenitor star that produces the sdB star or WD, the resultant binary should be quite compact (see Section~\ref{sec:formation}). As time increases after the CE phase, the binary orbit will shrink due to angular momentum losses, the minimum rate of which is from the emission of gravitational radiation (see e.g., Landau \& Lifshitz 1962; Rappaport et al.~1983). At some point, the low-mass companion star will fill its Roche lobe and begin to transfer mass to the WD.  At this point the system becomes an active cataclysmic variable and the pre-CV phase is over; we will address the further evolution of these systems in Section~\ref{sec:subeq_evol}.

We wish to answer the following question about these pre-CV systems - what is the shortest period that they can attain given the mass, radius, and evolutionary state of the companion star (taken to be a BD in this work) before it fills its Roche lobe?  We start by writing down an expression for the size of the Roche lobe, $R_{\rm L}$, as a function of the mass ratio, $q$, of the two stars and the orbital separation, $a$, assuming a circular orbit.  This takes the form
\begin{equation}
R_{\rm L} = f(q) \, a~,
\label{eqn:rL}
\end{equation}
where there are numerous functions in the literature to represent $f(q)$, some of which we discuss shortly.  To be clear, $4\pi R_{\rm L}^3/3$ is defined to be the volume of the Roche lobe, and we therefore refer to $R_{\rm L}$ as the `volumetric radius' of the Roche lobe.  If we insert this expression in Kepler's third law, we can write:
\begin{equation}
\frac{G (M_{\rm wd}+M_{\rm bd}) f^3(q)} {R_{\rm L}^3} = \left( \frac{2 \pi}{P} \right)^2
\label{eqn:kepler}
\end{equation}
where $M_{\rm wd}$ and $M_{\rm bd}$ are the masses of the WD and BD, respectively, $P$ is the orbital period, and $q \equiv M_{\rm bd}/M_{\rm wd}$.

A convenient analytic approximation to the volumetric radius of the Roche lobe, normalized to the orbital separation, was given by Kopal (1959):
\begin{equation}
f_{\rm K} = \frac{2}{3^{4/3}} \left( \frac{q}{1+q} \right)^{1/3},
\label{eqn:kopal}
\end{equation}
where the numerical value of the leading factor is 0.4622.  There is a more accurate expression derived by Eggleton (1983) based on an elegant fitting formula applied to the results of numerical integrations of the Roche-lobe volume:
\begin{equation}
f_{\rm E} = \frac{0.49 \, q^{2/3}}{0.6 \, q^{2/3}+\ln (1+q^{1/3})}.
\label{eqn:eggleton}
\end{equation}
For an extensive discussion of other evaluations of the Roche volume and fitting functions see Leahy \& Leahy (2015).

We now use the simpler, but more insightful, of the two expressions to derive the minimum period before Roche-lobe overflow commences, but then later show how that expression can be modified by the Eggleton (1983) expression to yield a more accurate result.  Inserting the expression $a=R_{\rm L}/f_{\rm K}(q)$ from Eqns.~(\ref{eqn:rL}) and (\ref{eqn:kopal}) into Eqn.~(\ref{eqn:kepler}), we find
\begin{equation}
P = \frac{2 \pi}{G^{1/2}} \sqrt{\frac{81}{8}} R_{\rm L}^{3/2} M_{\rm bd}^{-1/2} 
\end{equation}
which is completely independent of $M_{\rm wd}$, and this is the motivation behind using the Kopal (1959) formulation of $f_{\rm K}(q)$.  

\begin{figure}
\begin{center}
\includegraphics[width=\columnwidth]{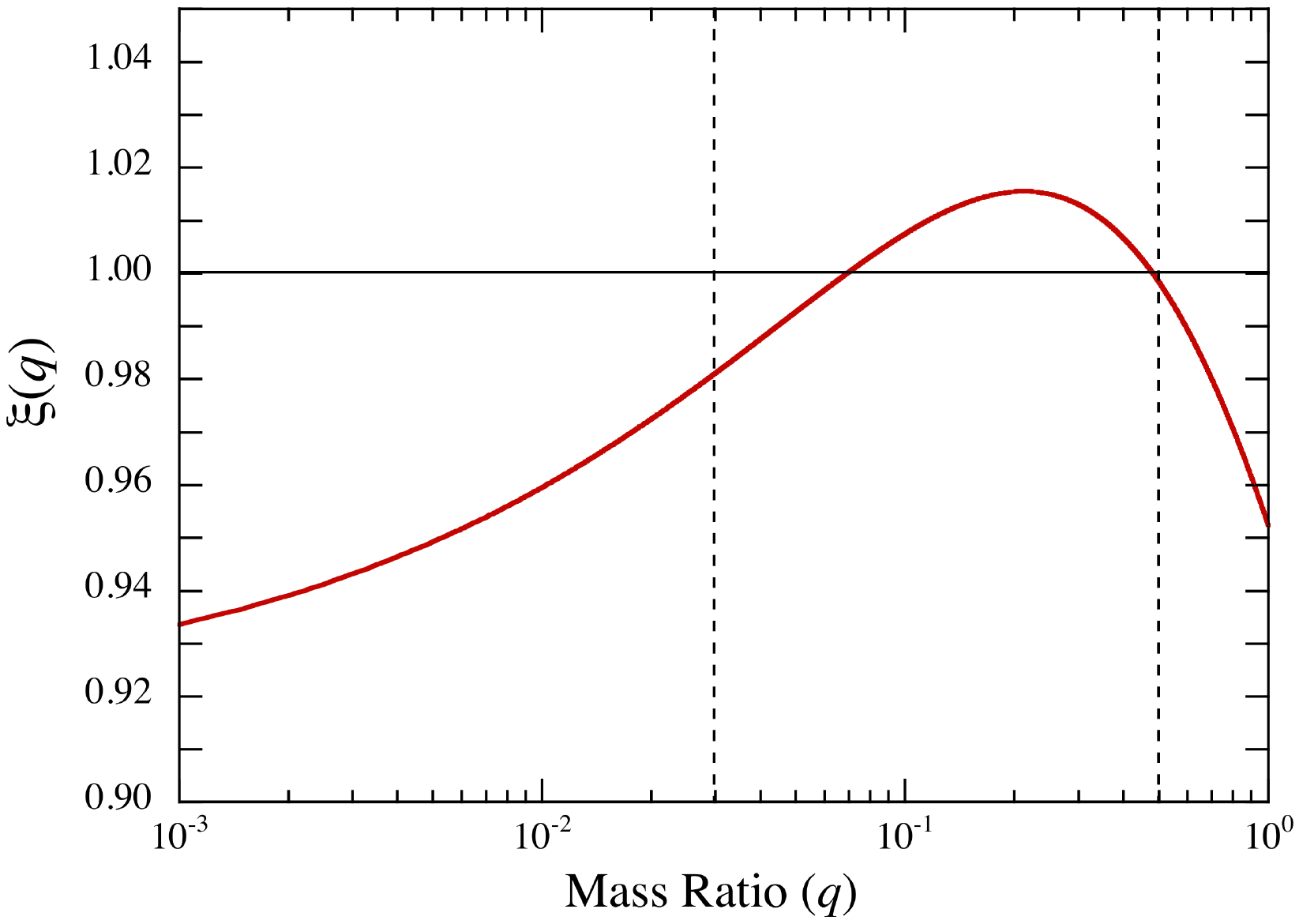}
\caption{Plot of the function $\xi(q) \equiv [f_{\rm K}(q)/f_{\rm E}(q)]^{3/2}$ (see Eqn.~(\ref{eqn:xi})). The two dashed vertical lines mark the approximate range of mass ratios of greatest interest in this work.  We see that $\xi(q)$ is unity over this range to within 2\%.}
\label{fig:xi} % Figure 3
\end{center}
\end{figure}

The minimum orbital period will come when the orbit shrinks to the point where the stellar radius equals $R_{\rm L}$, in which case we have
\begin{eqnarray}
P_{\rm min} & = &\frac{2 \pi}{G^{1/2}} \sqrt{\frac{81}{8}}R_{\rm bd}^{3/2} M_{\rm bd}^{-1/2} \nonumber \\
& = &  8.85 \left(\frac{R_{\rm bd}}{R_\odot}\right)^{3/2} \left(\frac{M_\odot}{M_{\rm bd}}\right)^{1/2} ~{\rm hr}
\label{eqn:Pmin1}
\end{eqnarray}
We now wish to utilize the more accurate $f_{\rm E}(q)$ expression for the Roche lobe dependence, while still casting the expression explicitly as a function only of $R_{\rm bd}$ and $M_{\rm bd}$, multiplied by a correction factor that is a very weakly dependent function of $q$. To accomplish that, we write 
\begin{eqnarray}
P_{\rm min}  & = & 8.85 \, \xi(q) \left(\frac{R_{\rm bd}}{R_\odot}\right)^{3/2} \left(\frac{M_\odot}{M_{\rm bd}}\right)^{1/2} ~{\rm hr} \nonumber \\
{\rm with} ~~\xi(q) & \equiv & \left[\frac{f_{\rm K}(q)}{f_{\rm E}(q)} \right]^{3/2}
\label{eqn:Pmin2}
\end{eqnarray}
The slowly varying function $\xi(q)$ is given explicitly by
\begin{equation}
\xi(q) = 0.916 \,\frac{[0.6 \, q^{2/3}+\ln (1+q^{1/3})]^{3/2}}{\sqrt{q (1+q)}}
\label{eqn:xi}
\end{equation}
We show a plot of $\xi(q)$ in Fig.~\ref{fig:xi}.  Over the range of greatest interest to us in this work, namely $0.03 \lesssim q \lesssim 0.5$, we find that $\xi(q)$ is limited to the range $1\pm 0.02$.  In the rest of this work we utilize Eqn.~(\ref{eqn:Pmin1}), and ignore any slight explicit dependence on the mass of the WD through $q$.  Moreover, it should be noted that the radii of these low-mass stars/BDs are not known to better than a few percent due largely to uncertainties in the input physics (e.g., EOS, opacities, nuclear cross-sections, diffusion) used to model the low-mass objects (Dorman et al. 1989; Tognelli et al. 2018).

\begin{figure*}
\begin{center}
\includegraphics[width=0.7\textwidth]{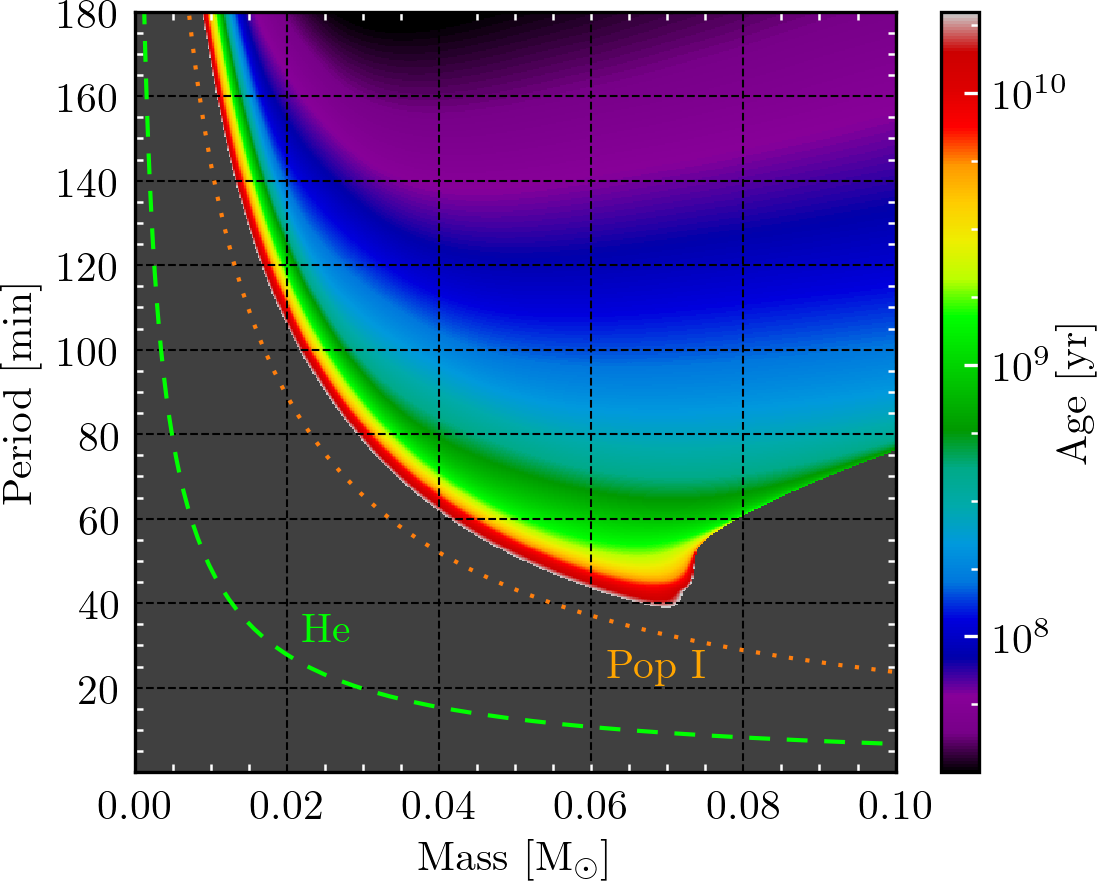}
\caption{The plane of minimum allowed orbital period vs.~BD mass.  The diagram is color-coded according to the logarithm of the cooling age of the BD since its birth.  The dark grey background indicates the region where there are no models.  For reference, we show two ``zero temperature'' models for the indicated compositions.}
\label{fig:pmin} % Figure 4
\end{center}
\end{figure*}

\begin{figure*}
\begin{center}
\includegraphics[width=0.7\textwidth]{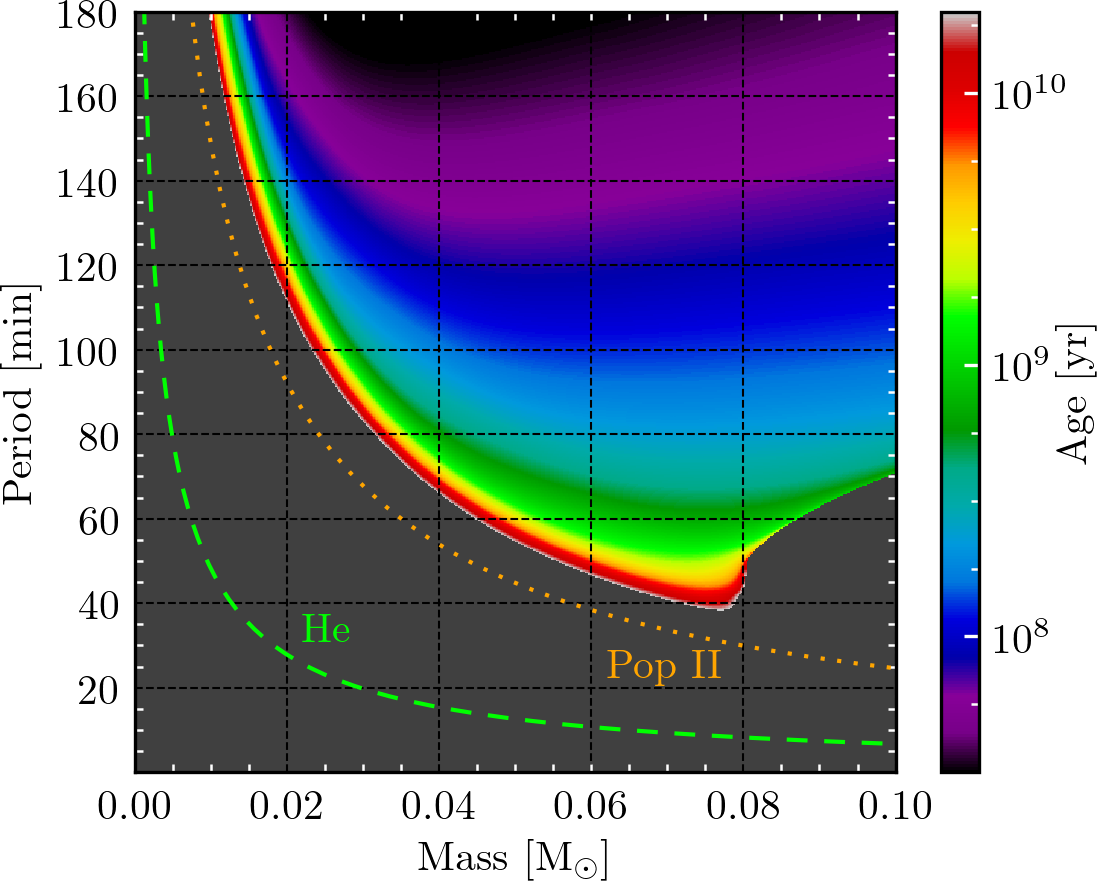}
\caption{Same as Figure \ref{fig:pmin} except for low metallicity, Z = 0.0001.}
\label{fig:pmin_lowZ} % Figure 5
\end{center}
\end{figure*}

\section{Brown Dwarf Cooling Models}
\label{sec:MESA}

In order to evaluate Eqn.~\eqref{eqn:Pmin1} we require BD cooling models for all masses of interest.  To this end we make use of Modules for Experiments in Stellar Astrophysics (MESA; Paxton et al. 2011, 2013, 2015, 2018) revision 10000.  The application of MESA to BDs and low-mass stars was presented in Paxton et al. (2013); we base our approach on the description therein and on the corresponding test suite case \texttt{make\_planets}.

We evolve close to 500 solar composition (Z = 0.02) models ranging from 0.002 to 0.1 M$_\odot$ in steps of $0.0002$ M$_\odot$.  We construct our starting models using the \texttt{create\_initial\_model} option (see Paxton et al. 2013, section 2.1).  This requires an initial mass $M$ and radius $R$, which we set as $R \simeq 5 M^{1/2}$, where $M$ and $R$ are both in solar units and are then converted to cgs units.  This choice for the radius is guided by the evolutionary results of Nelson et al. (1986, 1993) for contracting BDs on the Hayashi track\footnote{Because nuclear burning is negligible during this phase and assuming that the BD contracts at approximately constant $T_{\rm eff}$, then the radius of the BD (for a fixed age) should be $\propto M^{2/3}$ rather than $M^{1/2}$. However, the value of $T_{\rm eff}$ on the Hayashi track does have a very weak dependence on mass and this accounts for the difference in exponents.}.  We used the low temperature opacities of Freedman (2008) by setting \texttt{kappa\_lowT\_prefix} to \texttt{"lowT\_Freedman11"}.  Following the 
\texttt{make\_planets} test case, we set the controls  \texttt{max\_resid\_jump\_limit} and \texttt{max\_corr\_jump\_limit} to $10^{12}$.  This departure from the default values was necessitated so that MESA could avoid numerical difficulties while evolving models with masses of $\la 0.04$ M$_\odot$ at early ($\lesssim 1$ Myr) ages.  Each track was terminated at an age of 20 Gyr using the \texttt{max\_age} control.  For comparison purposes, we also ran another equivalent set of nearly 500 BD/MS cooling models for the low metallicity case of $Z = 0.0001$. 

The MESA models appear to reproduce the evolution of MS stars ($\gtrsim 0.075 M_\odot$) with a high degree of fidelity when compared to recently published results (Baraffe et al.~1998; 2015).  Except at very early ages ($\lesssim 10^8$ yr) where any discrepancy has no significant effect on our analysis, the radii of the MESA models match those of Baraffe et al. to within approximately 1-2\%.  We also compared the MESA models against our own low-mass BD evolution code and found similar good agreement at the 2-3\% level.  Our code employs the Lagrangian-based Henyey method and has been described in several papers (see, e.g., Nelson et al. 2004; Rappaport et al. 2017) as well as having been extensively tested (Maisonneuve 2007; Goliasch \& Nelson, 2015).   For lower masses, the `transition region' computed by the MESA code is about $0.002 M_\odot$ higher in mass than for either our models or those of Baraffe et al.  In the BD regime, when the MESA models have evolved to an age of $10^{10}$ yr (i.e., have reached $P_{\rm min}$), we find that they are generally smaller in radius by up to $\approx 5$\% in comparison to our own models.  This trend can also be seen when comparison is made to the Baraffe et al.~models but those tracks extend only to ages of 3 Gyr for masses $\leq 0.05 M_\odot$ (longer for the higher-mass tracks).  Thus a direct comparison is limited.  Nonetheless, we can conclude that the values of $P_{\rm min}$ derived from the MESA code are reasonable but that they should probably be regarded as a lower limit.

\section{Minimum Periods vs. Brown Dwarf Mass and Age}
\label{sec:pmin2}

We have used the BD cooling models discussed in Section~\ref{sec:MESA} to evaluate Eqn.~(\ref{eqn:Pmin1}) in order to compute the minimum periods of these pre-CV systems.  In Fig.~\ref{fig:pmin}, we show the results in the $P_{\rm min}-M$ plane where the color shading represents the BD evolutionary age. 

As expected from Eqn.~(\ref{fig:pmin}), the shortest allowed orbital period occurs for the highest mass BD at the oldest evolutionary ages (where $R_{\rm bd}$ is a minimum).  As can be seen from Fig.~\ref{fig:pmin} this occurs near a mass of $\simeq 0.072 M_\odot$.  The minimum orbital period for such BDs is 40 minutes at an age equal to a Hubble time.  This then is the absolute minimum period allowed for these pre-CV binaries.  For an arbitrary mass in the range of 0.01 to $0.072 M_\odot$, the minimum period can be fit to about 3\% by the following formula:
\begin{eqnarray}
%P_{\rm min} & = & 40 \,  \left [ 1.462 \left ({\frac {M}{M_{bd,0}}}\right )^{-\frac{1}{3}}  - 0.772 \left ({\frac {M}{M_{bd,0}}}\right )^{\frac{2}{3}} \right . \nonumber \\
%& & \left . + 0.308 \left ({\frac {M}{M_{bd,0}}}\right )^{\frac{5}{3}}  \right ]^{3/2} \quad {\rm min .}  \nonumber \\
P_{\rm min} \simeq  40  \, \left ({\frac {M_{\rm bd}}{M_{\rm bd,0}}}\right )^{-0.744}  ~{\rm min} 
\label{eqn:Pmin}
\end{eqnarray}  
over the range of $0.01 \lesssim M_{\rm bd}/M_\odot \lesssim 0.072$, where $M_{\rm bd,0}$ is a reference brown dwarf mass of $0.072 \, M_\odot$. 

After a Hubble time, the BDs have cooled sufficiently that their interiors are fully electron degenerate and their radii (and thus  $P_{\rm min}$) get progressively larger with decreasing mass.  Because the degeneracy is non-relativistic, the radius of these models is proportional to $M^{-1/3}$ (as would be expected for an $n = 3/2$ polytrope){\footnote{This progression continues until $M \approx 0.002 M_\odot$ ($\simeq 2 M_{\rm Jup}$) at which point the radius decreases due to contributions to the EOS from correlation energies and other atomic-based interactions.}}.  Stars with masses between $\simeq 0.072 M_\odot$ and $\simeq 0.075 M_\odot$ (`transition objects') are nearly in thermal equilibrium after a Hubble time (i.e., a non-negligible fraction of the luminosity radiated from the star's photosphere is generated by nuclear fusion. Because they have so much thermal energy their radii are substantially larger than would be expected for a completely degenerate configuration (see Fig.~\ref{fig:pmin}).  Higher-mass objects are able to contract and settle on the Main Sequence (thermal equilibrium) within a Hubble time.  The radii of stars near the end of the main sequence increase monotonically with increasing mass, as does $P_{\rm min}$.  

For reference, we have superposed several ``zero temperature'' models for different chemically-homogeneous  compositions (Pop I, Pop II, and pure He) in Figs.~\ref{fig:pmin} and \ref{fig:pmin_lowZ}.  These were calculated using the Zapolsky \& Salpeter (1969) equation of state (EOS).  For a given chemical composition and mass, these curves represent the absolute limit for $P_{\rm min}$ (corresponding to the smallest possible radius for the object).
The dotted Pop I curve illustrates that even after a Hubble time all of the BDs still have some residual thermal energy  that would be eventually radiated away.  If the star were to be composed of pure He, then it would be theoretically possible for the binary to evolve to very short orbital periods (see, e.g., Nelson et al. 1986).  Values of $P_{\rm min}$ as short as $\simeq 6$ minutes are possible and such ultra-short period systems have been observed as members of the AM CVn class of binaries.  

In Fig.~\ref{fig:pmin_lowZ} we show the same type of plot for $P_{\rm min}$ vs.~$M$ but for extremely metal-poor ($Z=0.0001$) Population II stars (the color contours denote the cooling ages).  There are several notable differences between the two figures: (i) the end of the MS is displaced to a larger mass ($\simeq 0.081 M_\odot$) and the highest BD mass (i.e., below the `transition region') is $\simeq 0.078 \, M_\odot$; and, (ii) the radii of Pop. II BDs are slightly larger than the corresponding Pop. I BDs (corresponding to a higher $P_{\rm min}$) after a Hubble time has elapsed.

\section{Subsequent Evolution After Contact}
\label{sec:subeq_evol}  

Eventually, the systems being discussed in this work will shrink under the influence of gravitational wave radiation (assuming that magnetic braking plays no significant role; Rappaport et al. 1983) to a state where the low-mass pre-CV companions begin to overflow their Roche lobes.  They can do so at periods $\approx 40-80$ minutes (so long as their radii are consistent with models having cooling ages $\gtrsim$ Gyr).  Therefore, they can produce CVs below $P_{\rm CV,min}$ (e.g., Politano 2004), and such a system might be detected as a faint dwarf nova with  very long intervals between outbursts (Howell et al. 1997).  We describe very briefly here how the system subsequently evolves.

Faulkner (1971) derives an ordinary differential equation describing the evolution of Roche-lobe overflowing binaries where the donor star has a power-law mass-radius relation, $R \propto M^n$.  The expression for the evolution of the mass fraction, $\mu$ of the donor star is given by:
\begin{equation}
\frac{d \tau}{d\mu}  =  \frac{\left[\mu -(1- \mu) ((n-\beta)/2+1)\right]}{(1-\mu)^2\, \mu^{2-4(n-\beta)}} 
\end{equation}
where $\mu = M_{\rm bd}/(M_{\rm bd}+M_{\rm wd})$ and $\beta$ is the power-law index for the scaling of the Roche lobe radius ($R_L/a \propto \mu^\beta$).  The variable $\tau$ is the scaled time ($\tau = t/T$) which we discuss below.  For the problem at hand, the mass-radius relation of a cold (degenerate) BD is $n \approx -1/3$ and Eqn.~(\ref{eqn:kopal}) requires that $\beta = 1/3$.  With these values, the differential equation for the evolution reduces to
\begin{equation}
\frac{d \tau}{d\mu}  = \frac{(5 \mu/3 -2/3)}{(1-\mu)^2\, \mu^{14/3}}. 
\label{eqn:mudot}
\end{equation}
% This integral is analytic, but not particularly insightful.
For small $\mu$, as is the case in these systems, we can neglect $\mu$ compared to terms of order unity, and solve for $\mu(t)$ and $P(t)$:
\begin{equation}
\mu \simeq \mu_0 \left[1+\frac{11}{2} \mu_0^{11/3} \frac{t}{T} \right]^{-3/11} \equiv \mu_0 \left[1+ \frac{t}{\mathcal{T}} \right]^{-3/11}
\label{eqn:muoft}
\end{equation}
\begin{equation}
P \simeq P_0 \left[1+\frac{11}{2} \mu_0^{11/3} \frac{t}{T} \right]^{3/11} \equiv P_0 \left[1+ \frac{t}{\mathcal{T}} \right]^{3/11}
\end{equation}
where $\mu_0$ is the mass fraction when mass transfer commences (at $t=0$), and where we have absorbed the extra constants into the definition of $\mathcal{T}$ (more general solutions for non-conservative mass transfer can be found in Chau \& Nelson 1983).  Now we can simply invert Eqn.~(\ref{eqn:mudot}) and utilize the result of Eqn.~(\ref{eqn:muoft}) to find:
\begin{equation}
\frac{dM_{\rm bd}}{dt} \simeq -\frac{3\mu_0}{2} \mu^{14/3} \frac{M_{\rm wd}}{T} \simeq -\frac{3\mu_0}{11} \left[1+\frac{t}{\mathcal{T}}\right]^{-14/11} \frac{M_{\rm wd}}{\mathcal{T}}.
\end{equation}
Finally, we define the characteristic timescale, $\mathcal{T}$ which governs the evolution of the BD mass fraction, $\mu$, the orbital period, $P$, and the mass transfer rate, $\dot M_{\rm bd}$:
\begin{equation}
\mathcal{T} = \frac{10 \,c^5 P_0^{8/3}}{352 \,(2\pi)^{8/3}G^{5/3}(M_{\rm bd}+M_{\rm wd})^{5/3} \mu_0}
\label{eqn:tau1}
\end{equation}
or, in more practical units
\begin{equation}
\mathcal{T} = 145 \left(\frac{P_0}{1\,{\rm hr}}\right)^{8/3} \left(\frac{M_\odot}{M_{\rm bd}+M_{\rm wd}}\right)^{5/3} \left(\frac{0.1}{\mu_0}\right) ~ {\rm Myr}
\label{eqn:tau2}
\end{equation}
Thus, the characteristic timescale for the binary, once Roche-lobe contact has been established, to increase its period from as short as 40 minutes back up to $\sim$70-80 minutes is of the order of a hundred Myr.  After this time, it would presumably resemble any normal CV that had evolved from a much longer period and an initially much more massive (but chemically unevolved) donor star.

\begin{figure}
\begin{center}
\includegraphics[width=\columnwidth]{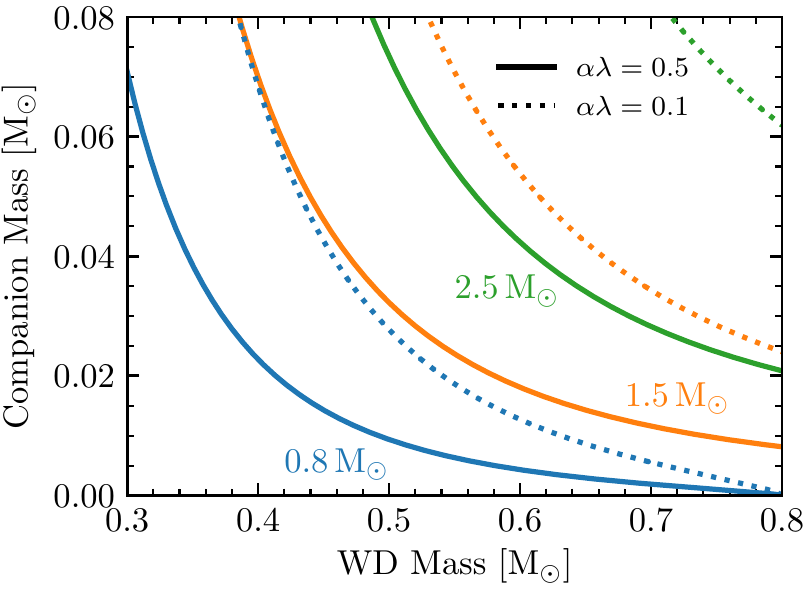}
\caption{The minimum companion mass, for any given WD mass, that is required to eject the common envelope of the WD progenitor star.  Curves are shown for three different representative masses of the progenitor: 0.8, 1.5 and 2.5\,$M_\odot$. The solid and dashed curves are for illustrative values for the CE parameter $\lambda \alpha = 0.5$ and 0.1, respectively. These curves are solutions that result from setting the right-hand sides of Eqn.~(\ref{eqn:WDBD}) and Eqn.~(\ref{eqn:Pmin}) equal to each other.}
\label{fig:Masses_CE}   % Figure 6
\end{center}
\end{figure}

\section{Minimum Companion Mass for CE Ejection}
\label{sec:minmass}  

We return here to try to answer the question in more detail of what is the lowest companion mass that can successfully eject the WD progenitor's envelope.  By `successful', we simply mean that at the end of the common envelope phase, the companion is still underfilling its Roche lobe.  To find the minimum such companion mass, we take the smallest radius for a given mass companion, which in turn, corresponds to the shortest periods found in Fig.~\ref{fig:pmin}.  We further consider only brown dwarf and planetary masses.  The locus of points near the bottom boundary of Fig.~\ref{fig:pmin} can be approximated to within $\approx 3$\% by Eqn.~(\ref{eqn:Pmin}). We can then equate the right-hand sides of Eqns.~(\ref{eqn:WDBD}) and (\ref{eqn:Pmin}) to find a relation for the minimum brown-dwarf mass (or planetary mass) required to eject the envelope of the white dwarf progenitor.
%\begin{equation}
%P_{\rm min} \simeq 0.004 \, M_{\rm comp}^{-0.744} ~{\rm days}  \quad  (0.008 \lesssim M_{\rm comp}/M_\odot \lesssim 0.072)
%\label{eqn:pminbound}
%\end{equation}

For any adopted $\lambda \alpha$ value, and assumed mass for the WD progenitor star, we can solve numerically for the mass of the companion star, $M_{\rm comp}$, as a function of the mass of the white dwarf.  Figure \ref{fig:Masses_CE} shows the minimum companion mass, for any given WD mass, that is required to eject the common envelope of the WD progenitor star for three different representative masses of the progenitor: 0.8, 1.5 and 2.5\,$M_\odot$.  We chose these three examples because if $M_p \lesssim 0.8 \,M_\odot$ the star would not have sufficient time to evolve to the state we are considering, while if $M_p \gtrsim 2.5 \, M_\odot$ the approximations that go into formulating Eqn.~(\ref{eqn:WDBD}) are no longer valid.  A value of $M_p \simeq 1.5 \, M_\odot$ might be considered `typical' based on population synthesis studies.  The solid and dashed curves are for illustrative values for the CE parameter $\lambda \alpha = 0.5$ and 0.1, respectively.

As can be seen from Fig.~\ref{fig:Masses_CE}, rather low-mass companions can indeed successfully eject the CE as long as the WD mass is sufficiently high. For stars with initial mass $\gtrsim 1\,M_\odot$, companions down to a few Jupiter masses can, in principle, eject the common envelope of the WD progenitor for WDs of mass $\lesssim 0.8 \, M_\odot$ (i.e., near the maximum value expected for the low-to-medium mass primaries we are considering).  Whether these planetary-mass objects can actually survive a common envelope is another question (see, e.g., Beuermann et al. 2010) and beyond the scope of this paper.

\section{Population Synthesis}
\label{sec:popsyn} 

In order to estimate the relative numbers of BD+WD systems that might be discovered with orbital periods in the range of 40-68 minutes and to understand how the results depend on assumptions concerning the formation of the systems, we have carried out a rudimentary binary population synthesis (BPS). BPS analyses of binaries containing BD/MS+WD stars (including CVs) have been continuously refined since the pioneering work of de Kool and Ritter (1993). Subsequent work has been carried out by Howell et al. (2001), Podsiadlowski et al. (2001), Willems and Kolb (2004), Nelson et al. (2004), Lu et al. (2006), Politano and Weiler (2007), Davis et al. (2010), Zorotovic et al. (2011), Goliasch and Nelson (2015), amongst others.

To carry out the analysis we invoke the $\alpha$-formalism which is concerned with energy conservation as described in Section~\ref{sec:formation}.  Our approach is similar to the one used by Howell et al. (2001) and updated by Goliasch and Nelson (2015). However, unlike their calculations, we assume that the initial mass distribution of the BD/MS (secondary) star is uncorrelated with the mass of the white dwarf progenitor (primary). The following list briefly describes the prescriptions that are incorporated in our Monte Carlo simulation:
\\
1. The primary mass, $M_p$, is chosen using Eggleton's (1993) random-variable representation of the Miller \& Scalo (1979) IMF:
\begin{equation}
M_p(\mathcal{R}) = \frac{0.19 \mathcal{R}}{(1-\mathcal{R})^{0.75}+0.032 (1-\mathcal{R})^{0.25}}
    \label{eqn:MS}
\end{equation}
where $\mathcal{R}$ is a linearly selected random number ($\mathcal{R}  \in  [0,1]$).  In our simulation the range of $\mathcal{R}$ is $0.909904 < \mathcal{R} < 0.996085$, which limits the primary mass to the range of  0.95 and 8 $M_\odot$.
\\
2. The secondary mass (a brown dwarf or star with a mass of $\leq 0.10 M_\odot$) is chosen using the piece-wise continuous IMF devised by Kroupa (2001). 
\begin{eqnarray}
%{\frac{d\mathcal{P}}{dM_s}}
\mathcal{P}(M_s) &=& C\ {M_s}^{-0.3}  \ , \quad (0.01 < M_s/M_\odot < 0.08) \nonumber \\
&=&  C \ 0.08 {M_s}^{-1.3} \ ,  \ (0.08 < M_s/M_\odot < 0.50) \nonumber \\
    \label{eqn:Kroupa}
\end{eqnarray}
where $\mathcal{P}$ is the probability density function and $C$ is a normalization constant. 
\\
3. The two masses are chosen independently; i.e., they are considered to be uncorrelated. \\
4. The initial orbital period of the primordial binary is chosen randomly according to a uniform distribution in $\log(P_i)$, where $P_i$, is defined by Eqn.~(\ref{eqn:PM}) and is limited to the range $10^{0.5}$ to $10^{6}$ days. \\
5. Based on the mass of the primary star (the WD/sdB progenitor) we use the time, $\tau_{\rm bgb}$ to reach the base of the giant branch given by the following prescription of Hurley et al.~(2002):
\begin{equation}
t_{\rm ev} = \frac{1594+2707m_p^4+147m_p^{5.5}+m_p^7}{0.04142m_p^2+0.3426m_p^7}  ~{\rm Myr}
\end{equation}
as a simple, single measure of how long it takes the primary to fill its Roche lobe.  
\\
6. Given the values for $P_i$, $M_p$, and $M_s$, we then solve for the mass of the degenerate core of the giant (primary) at the time that it fills its Roche lobe.  This mass is taken to be $M_{\rm wd}$.  We also checked to ensure that the mass transfer would be dynamically unstable (otherwise a CE would not be possible).  We tested several different values of the CE parameters $\alpha \lambda$, for values in the range of 0.01 to 1. 
\\
7. We then solve Eqn.~(\ref{eqn:WDBD}) for the orbital period immediately after the common envelope phase ($P_{\rm PCE}$). Note that that value of $\alpha \lambda$ is taken to be a constant, independent of the mass or evolutionary state of the primary star.  
\\
8. The birth-rate function (BRF) in the Galaxy is assumed to be constant and the age of the disk of the Galaxy is taken to be $t_{\rm gal}=10^{10}$yr.  We are then able to follow the orbital evolution over a time interval of up to $t_{\rm gal} - t_{\rm ev}$, with each instant in time within the interval corresponding to a different `birth time'.  Should the pre-CV evolve into contact (i.e., a semi-detached state) during the available time interval, its evolution is ignored from that point onwards.
\\ 

\begin{figure}
\begin{center}
\includegraphics[width=1.0 \columnwidth]{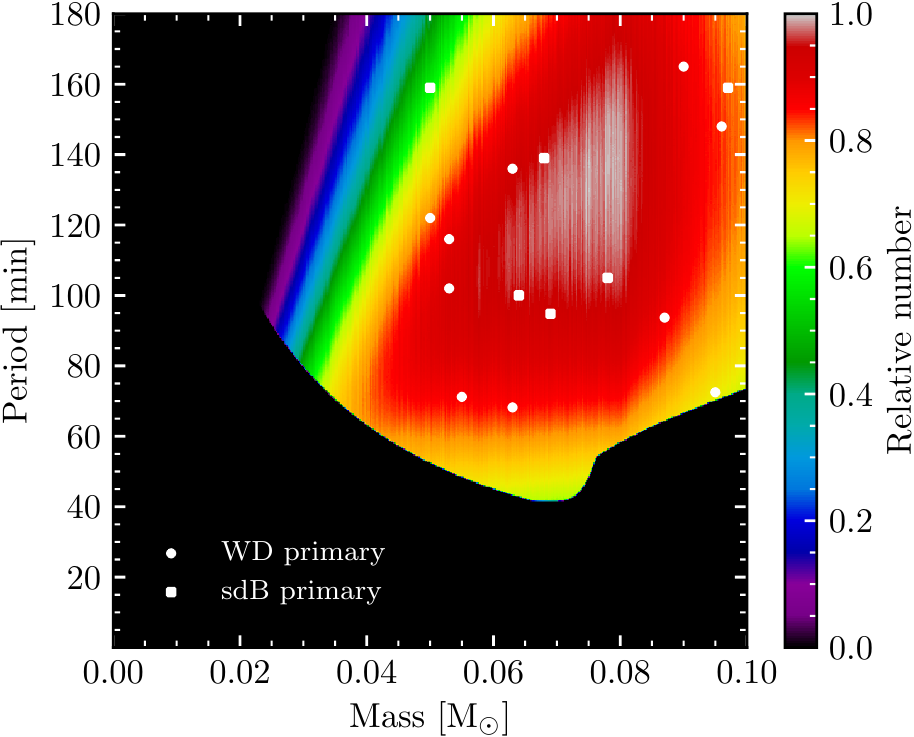}
\caption{Results of the population synthesis of BD/MS+WD binaries (Population I). The color coding is linearly scaled with the relative probability of finding systems in the $P_{\rm orb}-M$ plane at the current epoch.  See Sec.~\ref{sec:popsyn} for details of how the simulation was done.  The white squares and circles are known systems (see Table \ref{tbl:preCVs}).}
\label{fig:popsyn} % Figure 7
\end{center}
\end{figure}

For each BPS simulation we started by choosing $2 \times 10^9$ primordial binaries as described by the above set of prescriptions.  All of the `successful' pre-CVs were then evolved until either their age exceeded that of the Galaxy or they became CVs.   For the choice of $\alpha\lambda = 0.15$, approximately $5\times10^7$ systems succeed.  Figure~\ref{fig:popsyn} shows the relative probability of finding pre-CVs at the current epoch in the $P_{\rm orb}-M_s$ plane for this set of parameters.  Our choice of $\alpha\lambda = 0.15$ as the best representative case is somewhat guided by previous BPS results.  For example, Zorotovic et al. (2011) claim that a choice of $\alpha$ in the range of 0.2 to 0.3 leads to the simultaneous solution for all PCEBs in their sample{\footnote{There are differences in the treatment of the physics (e.g., recombination) that make direct comparison impossible.  Also different formulations of the CE analysis lead to difficulties in making more detailed comparisons.}}.  In addition, Davis et al. (2010) claim that values of $\alpha \gtrsim 0.1$ can satisfactorily describe the observed distribution of PCEBs with low-mass secondary stars.  Similar claims were made by Goliasch \& Nelson (2015).  Moreover, by carrying out BPS simulations over the range of $0.01 \leq \alpha\lambda \leq 1$, we were able to conclude that if the observations are not significantly biased then a value of 0.15 nicely reproduces the distribution of observed pre-CVs (see Fig.~~\ref{fig:popsyn}). 

Two things are apparent from a perusal of the figure.  First, if we look at the location of the known systems superposed on this plot, it seems qualitatively reasonable to believe that there exist BD+WD systems with periods between 40 and 68 minutes that simply have not yet been found.  More quantitatively, we can show from the simulated data that only about 11\% of all the systems with masses between 0.05 and $0.075\,M_\odot$ are expected to have periods in the range of 40-68 min.  Given that there are only about 10 systems that have been discovered in that mass range, we wouldn't necessarily expect that one of these ultra-short period systems should have been discovered yet.  If we assume that $\alpha\lambda \approx 0.25$, then this percentage (for the same mass range) becomes even smaller ($\approx 5$\%).  We also note that we were able to corroborate previous claims in the literature (e.g., Goliasch \& Nelson, 2015) that as $\alpha\lambda$ becomes progressively smaller, the number of successful systems monotonically decreases.  

Second, the known systems shown in Fig.~\ref{fig:popsyn} include none with masses $\lesssim 0.05\,M_\odot$.  This fits reasonably well with the BPS predictions for $\alpha\lambda = 0.15$.  However, there are a number of important caveats as to why such ultra-low mass secondaries have not been discovered.  The first concerns the physics associated with the CE efficiency ($\alpha$) and binding energy ($\lambda$). Both of these parameters are highly uncertain and, in the case of $\lambda$, it probably should not be treated as a constant (i.e., independent of the mass and evolutionary state of the WD progenitor).  The reason that the consistency for $\alpha\lambda = 0.15$ seems so good is that there is a ``mass-cutoff'' line that extends from about $0.025 M_\odot$ at 80 minutes to a mass and period of about $0.045 M_\odot$ and 180 minutes.  This cutoff effectively eliminates most masses of $\lesssim 0.05\,M_\odot$.  However, its location depends sensitively on the choice of $\alpha\lambda$.

The cutoff arises because systems with low-mass BDs cannot convert enough orbital energy as they spiral inwards to expel the envelope before eventually merging with the core of the giant during the CE phase.  Moreover, the lowest-mass BDs also have larger radii which exacerbates the situation (i.e., they can merge more easily).  It is very important to note that this cutoff can be shifted (more or less in a parallel line) to either higher or lower masses by changing $\alpha\lambda$.  For example, if $\alpha\lambda = 0.5$, the cutoff is pushed to masses as small as $\approx 0.015 M_\odot$.  Moreover, for $\alpha \lambda = 0.5$, the population synthesis predicts that the majority of BD+WD binaries should be comprised of BDs with masses of $\lesssim 0.04 M_\odot$ (contrary to what is  observed).  On the other hand, if $\alpha\lambda = 0.05$, then the mass cutoff precludes most systems that contain bona fide brown dwarfs (i.e., $M_{bd} \gtrsim 0.08 M_\odot$).

An approximate expression for this cutoff curve can be derived by taking Eqn.~(\ref{eqn:WDBD}) and maximizing it [i.e., $P_{\rm PCE}(M_{\rm bd},\alpha\lambda$)] for all possible values of $M_{\rm wd}$ and $ M_{p}$. Because of the strong dependence of $P_{\rm PCE}$ on $M_{\rm wd}$ (approximately the 8$^{\rm th}$ power), maximization occurs when $M_{\rm wd}$ is as large as possible given that $M_p$ should be minimized.  Based on various combinations of these parameters, $P_{\rm PCE}$ is maximized for a given $M_{\rm bd}$ when we set $M_{\rm wd} \simeq 0.6 M_\odot$ and $M_{p} \simeq 1.05 M_\odot$.  We also take $(M_{\rm bd}+M_{\rm wd}) \approx 0.65 \, M_\odot$ since $M_{\rm bd} \ll M_{\rm wd}$.  This yields
\begin{equation}
P_{\rm cutoff} \simeq 300\mbox{\, } {(\alpha\lambda)}^{3/2} \, \left( {\frac{M_{\rm bd} }{0.01M_{\odot } }} \right)^{3/2}\mbox{\, \, min.}
\label{eqn:cutoff}
\end{equation}
It is important to note that this equation only applies to BDs and very low-mass Pop. I MS stars.  

Thus the paucity of low-mass BDs ($\lesssim 0.05 M_\odot$) may be explained by mergers during the CE phase for appropriate values of $\alpha\lambda$.  The problem with this explanation is that it is necessarily heuristic because a precise quantification of the masses of the stellar components (and their binary correlation) is hard to infer and because the physics associated with the CE process is not well-understood.  Other possible reasons for the non-discovery of pre-CVs containing low-mass BDs include: (i) a sharp attenuation of the BD IMF at ultra-low masses; (ii) ablation/heating of the BD as it spirals in through the envelope of the WD progenitor; and, (iii) observational selection effects.  The first of these possibilities is quite plausible because the IMF of low-mass stars and especially BDs is not well known.  Several IMFs have been published based on either simple power laws (e.g., Salpeter, 1955), log-normal distributions (Miller \& Scalo 1979), or piece-wise continuous distributions (e.g., Kroupa, 2001; Chabrier, 2003).  But there remain very large uncertainties in the IMF for masses $\lesssim 0.05 M_\odot$.  To complicate matters, Thies \& Kroupa (2007, 2008) have suggested that because of a sharp decrease in the fraction of binaries containing BDs (related to the `BD desert'), it is likely that a two-component IMF is needed that takes into account the distinctly different modes of formation of (1) BDs and (2) low-mass stars ($\gtrsim 0.10 M_\odot$).  This obviously makes the estimation of the space density of very low-mass pre-CVs very difficult.

It may also be possible that very low-mass BDs experience serious ablation as they spiral through the common envelope.  The usual claim is that low-mass stars, such as late M dwarfs, are relatively pristine after completing the CE phase (see, e.g., Maxted et al. 1998).  But BDs with masses of about 0.01 to 0.02 $M_\odot$ have densities that are about an order of magnitude lower than very low-mass M dwarfs.  An investigation of the effects of the spiral-in are being undertaken by Turcotte et al. (2018) using the FLASH hydro code.  Finally we note that the lack of observed pre-CVs with very low-mass BDs may be due to observational selection effects.  This is in turn depends on the methods being employed to discover this class of pre-CVs.  A detailed analysis of this possibility is beyond the scope of our paper. 

One of the advantages of carrying out a population synthesis is that the space density of these low-period BD+WD binaries  can be approximately calculated.  Assuming a constant birtrate function (BRF) that is normalized to the production of 0.4 WDs/yr in the galactic disk, the predicted number density of these systems for $\alpha\lambda = 0.15$ is between $\approx 10^{-6}$ to  $3\times 10^{-6}$ pc$^{-3}$.  This is about an order of magnitude lower than that for CVs.  The reason for the range in the estimate is due to the assumed fraction of BDs that form in binary systems.  The lower end of the range assumes that about 15\% of BDs are born in binaries.  It should be noted that this estimate also depends sensitively on the choice of $\alpha\lambda$.  However, observations seem to constrain this parameter over a reasonably narrow range.

\section{Summary and Conclusions}
\label{sec:conclusion}

We present an analysis of pre-CV systems and their evolution.  We derive an analytic expression that relates the post-CE orbital period to the masses of the WD progenitor, its core mass, and the mass of the low-mass companion star. We also present analytic expressions for the minimum allowed orbital period of the pre-CV as a function of only the mass and radius of the low-mass companion.  Using a suite of MESA models, we find that the minimum orbital period for such a system is 40 minutes, corresponding to a BD with a mass of 0.07 $M_\odot$ and an age equal to a Hubble time.  For low metallicity models, the mass at which the minimum period occurs shifts up to near 0.08\,$M_\odot$, but the minimum period of 40 minutes remains the same.  The very existence of these short-period WD+BD objects provides information about CE evolution process involving low-mass secondaries.

We consider pre-CV systems with both WD and sdB primary stars.  For sdB stars that are He core burning, the timescale to come into contact due the emission of gravitational waves is longer than the nuclear timescale.  Thus, these objects will have evolved to become WD+BD binaries by the onset of mass transfer, and the observed systems will in fact spend most of their inspiral times as WD+BD binaries.  Such an evolution seems plausible even for the shortest period system (68 min; EPIC 212235321) among the known pre-CV WD+BD binaries (see Table \ref{tbl:preCVs} and Fig.~\ref{fig:contact}).

Once mass transfer does commence, these objects will begin to evolve towards longer periods.  We derive analytic expressions for the evolution of the mass, period, and mass-transfer rate during this phase.  For a timescale of order 100 Myr, these objects will be H mass transferring systems below the period minimum for conventional CVs.  Once they attain an orbital period of 70-80 minutes, they will closely resemble more conventional CVs that have descended from binaries containing much more massive companion (donor) stars.

We discuss the minimum companion mass, for any given WD mass, that is required to successfully eject the common envelope of the WD progenitor.  We find that for stars with initial mass between $\simeq 0.8\,M_\odot$ and 1\,$M_\odot$, companions down to a few Jupiter masses can plausibly eject the common envelope of the WD progenitor for the highest mass WDs we consider ($0.8 \, M_\odot$).  We leave it for other work to tackle the issue of whether these planetary-mass objects can actually survive a common envelope.

Finally, we have carried out a rudimentary population synthesis study of pre-CVs with low-mass stars (with a focus on BDs) to guide our expectations for where in the $P_{\rm orb}-M$ plane these systems should be found.  A reasonable match to the observations is obtained for the common-envelope parameter, $\alpha \lambda \approx 0.15$.  This could help explain the dearth of pre-CVs with $M_{\rm bd} \lesssim 0.05 \, M_\odot$.  It also allows us to estimate the space density of this class of pre-CVs in the solar neighborhood to be $\approx 10^{-6}$ pc$^{-3}$.

\acknowledgments

Support for this
work was provided by NASA through Hubble Fellowship to J.S. through grant \#
HST-HF2-51382.001-A awarded by the Space Telescope Science Institute,
which is operated by the Association of Universities for Research in
Astronomy, Inc., for NASA, under contract NAS5-26555.  This research
has made use of NASA's Astrophysics Data System.
L.N. thanks the Natural Sciences and Engineering Research Council (Canada) for financial support provided through a Discovery grant.  We also thank Calcul Qu\'ebec, the Canada Foundation for Innovation (CFI), NanoQu\'{e}bec, RMGA, and the Fonds de recherche du Qu\'{e}bec - Nature et technologies (FRQNT) for computational facilites.  Finally we thank J. Aiken for his technical assistance.



%%%%%%%%%%%%%%%%%%%%%%%%%%%%%%%%%%%%%%%%%%%%%%%%%%

\appendix

\section{Comparison of Pop I and Pop II Brown Dwarfs in the $P_{\rm min}-M$ Plane}
\label{BD_populations}

In Section \ref{sec:pmin2} we compared the $P_{\rm min}-M$ plane for Population I and extremely metal-poor ($Z=0.0001$) Population II stars in Figs.~\ref{fig:pmin} and \ref{fig:pmin_lowZ}.  There we noted several differences between the two figures: (i) for Pop II stars the end of the MS is displaced to a larger mass ($\simeq 0.081 M_\odot$) and the highest BD mass (i.e., below the `transition region') is $\simeq 0.078 \, M_\odot$; and, (ii) the radii of Pop. II BDs are slightly larger than the corresponding Pop. I BDs (corresponding to a higher $P_{\rm min}$) after a Hubble time has elapsed.

The {\it first} difference is well known and can be explained by the fact that fully convective Pop. II MS stars ($\lesssim 0.3 M_\odot$) have much lower surface opacities ($\kappa_s$) than Pop. I stars of the same mass.  The photospheric opacities tend to be dominated by H$^-$ interactions; since most of the free electrons are supplied by ionized metals (not hydrogen), metal-poor atmospheres have greatly diminished opacities.  The atmospheres in very low-mass stars and BDs exhibit little to no superadiabaticity (i.e., convective mixing lengths are unimportant), and the photospheres act like valves to release the energy generated in the interiors.  Because lower radiative opacities imply that the energy can escape more easily, Pop. II stars require an enhanced rate of nuclear energy generation compared to a Pop. I stars in order to sustain nuclear quasi-equilibrium.  Thus, compared to Pop. I stars at the bottom of the MS, Pop. II stars require a higher mass because nuclear energy generation depends sensitively on the central temperature which in turn is positively correlated with the mass{\footnote{According to the Virial Theorem, $T_c \propto M/R$.  Thus for a fixed mass, low-metallicity stars (on the subdwarf sequence) have a smaller radius and higher luminosity than their Pop. I counterparts.}.  

The {\it second} difference, namely the larger radii of old Pop II BDs, can be traced back to the opacities and, to a much lesser extent, the EOS.  Pop. II stars on the Hayashi track (HT) initially contract faster than Pop. I stars with the same mass and radius.  This can be understood in terms of the Virial Theorem which requires that the energy radiated away during the contraction (i.e., the luminosity $L$) be approximately 1/2 of the absolute change in the gravitational potential energy ($\propto M^2/R$) of the object.  This assumes that nuclear energy generation is unimportant and that the gas in the star is `perfect'.  Thus the rate of contraction is given by
\begin{equation}
\frac{dR}{dt} \ \propto \ \frac{R^4 T_{\rm eff}^4}{M^2} \ \propto \ \frac{T_{\rm eff}^4}{g^2}
\label{eqn:Rdot1}
\end{equation}
where $T_{\rm eff}$ is the effective temperature of the star and $g$ is the gravitational acceleration at its surface.  Pop. II stars on the HT tend to have higher $T_{\rm eff}$s in order to compensate for their lower opacities.  Thus they contract more quickly than their Pop. I counterparts (for the same mass and radius).  

However, once Pop. II stars leave the HT and begin the phase of degenerate cooling, they contract less quickly than Pop. I stars.  The reason can be seen in Eqn.~(\ref{eqn:Rdot1}).  For BDs with comparable values of $g$,  $dR/dt$ is strongly dependent on $T_{\rm eff}$. Because Pop. II BDs initially evolve faster (as they expend their gravothermal energy), they approach the degenerate sequence more quickly and thus their $T_{\rm eff}$s become lower (H$^-$ opacities are no longer dominant at cool $T_{\rm eff}$s).  The slower rate of contraction compared to Pop. I BDs eventually causes the radii of the Pop. I BDs to become smaller than those of Pop. II BDs}{\footnote{The fact that the absolute minimum period of Pop. II BDs is still very near 40 minutes (as it is for Pop. I) is largely fortuitous.  Even though Pop. II BDs have larger degenerate radii, they can have higher masses compared to Pop. I models and these two effects offset each other.  Consequently, the Pop. II model with the shortest $P_{\rm min}$ has a mass of $\approx 0.078 \, M_\odot$ as opposed to $\approx 0.071 \, M_\odot$ for Pop. I.}.  

The EOS also plays a minor role in causing this behavior.  In comparing Fig.~\ref{fig:pmin} and Fig.~\ref{fig:pmin_lowZ}, we find that for a given mass, the Pop II $T=0$ curve corresponds to a slightly but consistently larger value of $P_{\rm min}$ (and hence radius) compared to that for Pop. I.  As the BDs become very cold, electron degeneracy is the primary contributor to the  pressure EOS (thermal contributions from the ions and Coulombic interactions are much less important).  The specific pressure (i.e., per unit mass) thus depends critically on the number density of degenerate electrons.  Because both H and He will be largely ionized by pressure ionization throughout the BD interiors, it is clear that the electrons from hydrogen atoms will contribute much more to the specific electron degeneracy pressure than helium.  Thus Pop. II zero-temperature models have slightly larger radii (and thus $P_{\rm min}$s) because of their enhanced hydrogen abundances relative to those of Pop. I models.  This is what is observed in the two figures (with the helium zero-temperature models showing how small the $P_{\rm min}$s can become).

\end{document}